\documentstyle{ichep}
\begin{document}

\title{Semileptonic $B$ Meson Decays and Interfering Amplitudes$^{\ddag}$}

\author{K. Honscheid$^{\dag}$}
\affil{Department of Physics, The Ohio State University, Columbus
Ohio 43210, USA}
\author{K. Schubert, R. Waldi}
\affil{Institut f\"ur Kern- und Teilchenphysik, Technische Universit\"at
Dresden, D-01062 Dresden, Germany}

\abstract{
Consequences of the interference between external and internal spectator
amplitudes for the lifetimes and semileptonic decay fractions of $B^0$ and
$B^+$ mesons are discussed. Extrapolating from the constructive interference
observed in 11 exclusive hadronic $B$ decays we find an inclusive semileptonic
decay fraction of $(11.2 \pm 0.5 \pm 1.7)$\%, significantly closer to
the experimental results than previous predictions.
}

\twocolumn[\maketitle]

\fnm{1}{E-mail: kh@mps.ohio-state.edu}
\fnm{2}{Invited talk presented at the International Conference on High Energy
Physics, Glasgow 1994.}

\section{Introduction}
Although there has been significant progress in the
calculation of QCD corrections in the decays of heavy flavour mesons,
there are still some unsolved puzzles.
One of the most intriguing is the low semileptonic
decay fraction of $B$ mesons \cite{theory}.
Ignoring the small $b \to u$ fraction, the $b$ quark in the $B$ meson
decays to a charm quark and emits a virtual $W$ boson. This can transform
itself into a lepton neutrino or a quark anti-quark pair. Taking into
account the color factors and making some crude assumption about the quark
masses, we can determine the relative rate of these processes and find
an semileptonic decay fraction of approximately 15\%. To obtain a more precise
number we have to correct for hadronic effects due to the exchange of
gluons between the quark lines. This enhances the hadronic
rate with respect to the semileptonic rates resulting in ${\cal{B}}_{s.l.}
\approx 13-14$\%. Bigi \etal have recently
performed an evaluation of ${\cal{B}}(B \to X e \nu)$ based on the
$1/m_Q$ expansion method in QCD \cite{mq} and found that the theory
cannot accomodate a semileptonic branching fraction of $B$ mesons of
less than 12.5 \%.

Experimentally, the semileptonic decay fraction of $B$ mesons has been
determined by the ARGUS and CLEO collaborations and by the four
LEP experiments. ${\cal{B}}_{s.l.}$ is determined by integrating
over the measured lepton momentum spectrum. However,
models have to be used to remove the background from $b \to c \to s$ cascade
decays.
The model dependence can be significantly reduced by selecting
 $\Upsilon (4S)$ decays with two final state leptons. A high momentum lepton
tags this reaction as a $B\bar{B}$ event while the other lepton is
used to measure the lepton momentum spectrum in semileptonic $B$ decay.
Following this procedure, CLEO has contributed a paper
to this conference quoting a value of $(10.36 \pm 0.17 \pm 0.40)$\%
for the semileptonic $B$ decay rate \cite{cleosl}.
This is significantly below the lower
bound allowed by theory and hence we have a problem.

\section{Interfering Amplitudes in Hadronic $B$ Decays}

A solution to this problem would be a further enhancement of the hadronic
decay rate with respect to the semileptonic rate. Hadronic $B$ decays proceed
via external or internal spectator diagrams. While the two diagrams lead
to different final states in $B^0$ decays, both processes produce the same
final state in charged $B$ decays and hence the
corresponding amplitudes will interfere.
These two amplitudes combined with the factorization hypothesis form the
framework of spectator models such as the model by Bauer, Stech, and
Wirbel \cite{bsw}.
These models have been surprisingly successful in describing many
features of heavy meson decay. Destructive interference between the
internal and external spectator amplitude for
hadronic $D^+$ decays reproduces the observed
$D^0-D^+$ lifetime difference.
Bauer, Stech, and Wirbel describe the two amplitudes by phenomenological
parameters $a_1$ and $a_2$. The values of these parameters have to be
determined by experiments. Destructive interference as observed in $D^+$
decays is described by a relative minus sign between $a_1$ and $a_2$. The
theoretical interpretation of these parameters is
controversial \cite{mq} but it was generally expected that a similar but
less pronounced interference pattern would be found in $B$ decay.
It came as surprise when the CLEO collaboration reported constructive
interference in all exclusive hadronic $B^+$ decays observed so far
\cite{bigb}.
\begin{table}[htb]
\label{br_ave}
\Table{|c|c|c|c|c|}{
  decay&
 exp. average $[\%]$&
Neubert et al. \cite{neubert}\\
\hline
$B^+ \rightarrow \bar{D}^0 \pi^+$&
$0.45 \pm 0.04 $&
$0.265(a_1 + 1.230 a_2)^2$\\
$B^+ \rightarrow \bar{D}^0 \rho^+$&
$1.10 \pm 0.18 $&
$0.622(a_1 + 0.662 a_2)^2$\\
$B^+ \rightarrow \bar{D}^{*0} \pi^+$&
$0.51 \pm 0.08 $&
$0.255(a_1 + 1.292 a_2)^2$\\
$B^+ \rightarrow \bar{D}^{*0} \rho^+$&
$1.32 \pm 0.31 $&
$.70(a_1^2 + 1.49 a_1a_2 + .64 a_2^2)$\\
$B^+ \rightarrow \psi K^+$&
$0.106 \pm 0.015$&
$1.819\, a_2^2$\\
$B^+ \rightarrow \psi K^{*+}$&
$0.17 \pm 0.05$&
$2.932\, a_2^2$\\
\hline
$B^0 \rightarrow D^- \pi^+$&
$0.26 \pm 0.04$&
$0.264\, a_1^2$\\
$B^0 \rightarrow D^- \rho^+$&
$0.69 \pm 0.14$&
$0.621\, a_1^2$\\
$B^0 \rightarrow D^{*-} \pi^+$&
$0.29 \pm 0.04$&
$0.254\, a_1^2$\\
$B^0 \rightarrow D^{*-} \rho^+$&
$0.74 \pm 0.16$&
$0.702\, a_1^2$\\
$B^0 \rightarrow \psi K^0$&
$0.069 \pm 0.022$&
$1.817\, a_2^2$\\
$B^0 \rightarrow \psi K^{*0}$&
$0.146 \pm 0.029$&
$2.927\, a_2^2$\\
}\caption{Experimental averages and
theoretically predicted decay fractions
for hadronic $B$ decays,
assuming $|V_{cb}|^2\cdot \tau_{B} = 2.35\, 10^{-15}$s, and
$f_D = f_{D^*} = 220$ MeV}
\end{table}
Combining the experimental decay fractions
measured by ARGUS and CLEO \cite{bhp} results in the averages
listed in Table \ref{br_ave}. The partial rates are determined under the
assumption of equal decay fractions of the $\Upsilon (4S)$ to $B^+B^-$
and $B^0 \bar{B}^0$ pairs, i.e. $f^{+-}/f^{00} = 1$. This
quantity is not well measured experimentally; we assume in the following
$f^{+-}/f^{00} = 1.0 \pm 0.1$.
\begin{table}[htb]
\label{ratio}
\Table{|c|c|c|}{
 & exp. average& Neubert et al. \cite{neubert} \\
\hline
$R_1=\frac{\Gamma(B^+ \rightarrow \bar{D}^0\pi^+)}{\Gamma(B^0
\rightarrow D^+\pi^-)}$&
$1.71 \pm 0.38$&
$(1 + 1.23 a_2/a_1)^2 $\\
$R_2=\frac{\Gamma(B^+ \rightarrow \bar{D}^0\rho^+)}{\Gamma(B^0
\rightarrow D^+\rho^-)}$&
$1.60 \pm 0.46$&
$(1 + 0.66 a_2/a_1)^2 $\\
$R_3=\frac{\Gamma(B^+ \rightarrow \bar{D}^{*0}\pi^+)}{\Gamma(B^0
\rightarrow D^{*+}\pi^-)}$&
$1.79 \pm 0.39$&
$(1 + 1.29 a_2/a_1)^2 $\\
}\caption{Experimental results and theoretical predictions for ratios of
$B^+$ and $B^0$ decay rates, scaled to $f_{D(D^*)}= 220$ MeV}
\end{table}
The relative sign between $a_1$ and $a_2$ can be obtained from
$B^+ \rightarrow \bar{D}^0$ and
$B^+ \rightarrow \bar{D}^{*0}$
decays, which have contributions from both amplitudes.
A relative plus
sign between the $a_1$ and the $a_2$ amplitudes would give
$\Gamma(B^+ \rightarrow \bar{D}^{(*)0} \pi(\rho)^+ )/
\Gamma(B^0 \rightarrow D^{(*)-} \pi(\rho)^+) > 1$, while a minus sign
would correspond to ratios below 1.
The experimental results and
a model prediction for the decay ratios in the modes
$D \pi^-$, $D \rho^- $, and $ D^{*} \pi^-$
are given in Table \ref{ratio}.
They show a clear preference for the positive sign.
The theoretical prediction for the decay
$B^+ \rightarrow D^{*0} \rho^+$ is too uncertain
\cite{rieckert}
to include this mode in the determination of
$a_1$ and $a_2$.
Taking ratios of $B^+$ and $B^0$
decays eliminates the uncertainties due to $|V_{cb}|$ but leaves
those originating from $\tau (B^+)/\tau (B^0)$ and
$f^{+-}/f^{00}$. The main difference between different models are details
of the $B\rightarrow \pi$ and $B \rightarrow \rho $ form factors.
The predictions also depend on the $D$ and $D^*$ decay
constants $f_D$ and $f_{D^*}$. Following Neubert et al.\ \cite{neubert}
we assume $f_D = f_{D^*}\, = \,220$ MeV.
On the experimental side, the error due to the $D^0$ decay fractions
cancels in the ratios involving $B \rightarrow D^*$ decays.
A least square fit with seven $B \to D^{(*)}$ modes from Table \ref{br_ave},
excluding only $B^+ \to D^{*0} \rho^+$, gives
$a_1 \, = \, 1.04 \pm 0.05$ and
$a_2 \, = \, 0.24 \pm 0.06$.

\section{Assumptions}

The distinction between interfering amplitudes for the $B^+$
and non-interfering for the $B^0$ may only be valid for
two-body decays.
On the other hand,
many-body final states will most likely start as two colour
singlet quark antiquark pairs, including
intermediate massive resonances.
Interference between final states via different resonant
channels involves strong phases
which modify the rate for each individual final state
in a random way and disappear in the sum of all states.
It seems therefore reasonable to extend the model for exclusive
two body decays to
the majority of hadronic final states in an inclusive picture
at the quark level. We assume that the formation of two
colour singlets is the essential step of hadron production,
which is
taken into account quantitatively by $a_1$ and $a_2$.
We neglect modifications by
decays into baryon anti-baryon pairs, where our assumption is
not valid.
\begin{table}
\Table{|c|c||c|c||c|c|}{
$B^+ (\bar{b} u ) $ & QCD &
 $ B^0(\bar{b} d ) $ & QCD & CKM & PS \\
 $\to $ &           &
 $\to $ &            &            &           \\
\hline
$ \bar{c} u\,e\nu $ & 0.86  &
$  \bar{c} d\,e\nu $ & 0.86  & & 1.00 \\
$ \bar{c} u\,\mu\nu $ & 0.86  &
$  \bar{c} d\,\mu\nu $ & 0.86 & & 0.99 \\
$ \bar{c} u\,\tau\nu $ & 0.86  &
$ \bar{c} d\,\tau\nu $ & 0.86  & & 0.23  \\
$ \bar{c} u\ \bar{d} u $ & $3(a_1+a_2)^2$ &
$ \bar{c} d\ \bar{d} u $ & $3a_1^2 $& $|V_{ud}|^2$&1.00 \\
                 &                 &
$  \bar{c} u\ \bar{d} d $ & $3a_2^2 $ & $ |V_{ud}|^2$ &1.00 \\
$ \bar{c} u\ \bar{s} u $  & $3(a_1+a_2)^2$ &
$ \bar{c} d\ \bar{s} u $ & $3a_1^2 $& $ |V_{us}|^2$&0.98 \\
        &     &
$  \bar{c} u\ \bar{s} d $ & $3a_2^2 $& $ |V_{us}|^2$&0.98 \\
$ \bar{c} u\ \bar{s} c $ & $3a_1^2 $&
$ \bar{c} d\ \bar{s} c $ &  $3a_1^2 $&$ |V_{cs}|^2    $  &0.48 \\
$ \bar{c} c\ \bar{s} u $ & $3a_2^2 $&
$ \bar{c} c\ \bar{s} d $ &  $3a_2^2 $&$ |V_{cs}|^2    $  &0.48 \\
$ \bar{c} u\ \bar{d} c $ &  $3a_1^2$&
$ \bar{c} d\ \bar{d} c $ &  $3a_1^2 $&$ |V_{cd}|^2    $ &0.49 \\
$ \bar{c} c\ \bar{d} u $ &  $3a_2^2 $&
$ \bar{c} c\ \bar{d} d $ &  $3a_2^2 $&$ |V_{cd}|^2    $ &0.49 \\
}
\caption{Contributions from all $b\to c$ spectator diagrams.
Partial widths are obtained as $\Gamma =
\Gamma_0(b \to c e^- \bar{\nu})\times
CKM\times QCD \times PS$.}
\label{result}
\end{table}
Under the assumption of duality, the coefficients $a_1$ and $a_2$ can be used
to predict the hadronic and semileptonic widths of the $B^+$ and $B^0$ mesons.
The individual contributions are listed in Table \ref{result}. $PS$ denotes the
relative phase space factor and the perturbative QCD correction for the
semileptonic width is given by \cite{cabb}
$$
\Gamma(b \to c e^- \bar{\nu}) \, = \, \Gamma_0
(1- \frac{2\pi}{3}\alpha_s + \frac{25}{6\pi}\alpha_s)
\approx 0.86\Gamma_0
$$
{}From the factors in Table \ref{result} we obtain
the following total widths, normalized
to the lowest order semileptonic width
$\Gamma_0(b \to c e^- \nu)$
$$\eqalign{
{\Gamma(B^+) /\Gamma_0} &=
 1.91 + 4.44 (a_1^2 + a_2^2) + 5.99 a_1 a_2\,, \cr
{\Gamma(B^0) /\Gamma_0} &=
 1.91 + 4.44 (a_1^2 + a_2^2)\,. \cr
}$$
Using these widths, we can calculate two important quantities.
\begin{itemize}
\item The average
semileptonic decay fraction of $B^0$ and $B^+$,
$$
{\cal{B}}(B \to e \nu X) =
{\frac{1}{ 2.22 + 5.16 (a_1^2 + a_2^2) + 3.49 a_1 a_2}}\,,
$$
decreases if $a_2$ changes sign from negative to positive.
\item The lifetime ratio
\begin{equation}
\tau(B^+) / \tau(B^0) =
 1 - {\frac{a_1 a_2}
 {0.32 + a_1 a_2 + 0.74 (a_1^2 + a_2^2)}}
\label{eq_life}
\end{equation}
is larger than 1 for negative and
smaller than 1 for positive values of $a_2$.
\end{itemize}
To give consistent results, we determine $a_1$ and $a_2$
in a fit to the hadronic decay fractions used above,
replacing the assumption of equal $B^+$ and $B^0$ lifetimes
with the inclusive prediction in
eq.~\ref{eq_life} to rescale the theoretical expectations
for $B^+$ and $B^0$ decays individually.
This fit gives $\chi^2 = 11.6$ for 8 degrees of freedom, and
$$\eqalign{
a_1 &= 1.05 \pm 0.03 \pm 0.10\cr
a_2 &= 0.227 \pm 0.012 \pm 0.022\cr
}$$
which implies
$$\eqalign{
{\cal{B}}(B \to e \nu X) &= (11.2\pm0.5\pm1.7)\% \cr
\tau(B^+) / \tau(B^0) &= 0.83\pm0.01\pm0.01\,,\cr
}$$
where the first error is statistical including uncertainties in
the $D^0$ and $D^+$ decay fractions, and
the second is from the
error on $V_{cb} \sqrt{\tau (B)}$. The uncertainty
on $f^{+-}/f^{00}$ yields a negligible error.
The predicted lifetime ratio is low but not
inconsistent with the current
experimental average of $1.00 \pm 0.07$ \cite{sharma}.
The semileptonic decay fraction is further reduced if we assume
a small contribution of penguin decays.
Assuming this fraction to be $2.5\%$ leads to $\chi^2 = 11.3$ and
${\cal{B}}(B \to e \nu X) = 10.9\%$, while all errors and the values of
$a_1$, $a_2$ and
$\tau(B^+) / \tau(B^0)$ remain essentially unchanged.

\section{Discussion}
The discrepancy between the theoretical
and the experimental semileptonic decay fraction of $B$ mesons can be
considerably reduced by the interpretation of recent results
on hadronic $B$ decays in the framework of a spectator model with interfering
amplitudes. Our basic assumption is that the constructive interference
observed in a few exclusive hadronic $B$ decays is a general feature of
$B$ mesons that can be described by two coefficients $a_1$ and $a_2$.
There is some experimental evidence that this assumption is correct
\begin{itemize}
\item The coefficient $a_2$ extracted from the interference observed
in $B^+ \to D^{(*)}$ decays agrees well with the $a_2$ value obtained
from $B$ to charmonium transitions that
can only proceed through the internal spectator diagram.
\item A QCD based calculation \cite{lepage}
 of inclusive $\psi$ production in $B$ decay
falls short of a recent CLEO measurement \cite{cleopsi}. However, if the
coefficients in the calculation are replaced by the measured values of $a_1$
and $a_2$ we find good agreement.
\item The measured $B \to \chi_{c1}$ decay fraction has been used to predict
the
$B \to \chi_{c2}$ rate \cite{lepage,cleopsi}.
Again the agreement with recent experimental
results can be improved by using $a_1$ and $a_2$ instead
of the QCD coefficients in the calculation.
\end{itemize}
A careful study of inclusive decays as well as a search for
color suppressed decays like $B^0 \to D^0 \pi^0$ will allow us
in the not too distant future to determine,
if $a_1$ and $a_2$ are really universal
coefficients in $B$ decays.


\Bibliography{99}
\bibitem{theory}%
G. Altarelli and S. Petrarca, PL B261, 303 (1991);
I. I. Bigi et al., PL B293,
430 (1992) and erratum ibid. B297, 477 (1993);
W. Palmer, B. Stech, PR D48, 4174 (1993).
\bibitem{cleosl} CLEO Collaboration, International Conference on HEP,
Glasgow, CLEO-CONF-94-6 (GLS 0243).
\bibitem{mq}
I. I. Bigi, B. Blok, M. A. Shifman, N. G.  Uraltsev, A. I. Vainshtein,
``B Decays'', ed. S. Stone, World Scientific (1994).
\bibitem{bigb}M. S. Alam et al., Phys. Rev. D50, 43 (1994).
\bibitem{bhp}
T. E. Browder, K. Honscheid, S. Playfer,
``B Decays'', ed. S. Stone, World Scientific (1994).
\bibitem{bsw}M. Bauer, B. Stech, M. Wirbel, ZP C34,103(1987).
\bibitem{neubert}M. Neubert, V. Rieckert, B. Stech in `Heavy Flavors',
ed.\ by A. J. Buras and M. Lindner,
World Scientific 1992.
\bibitem{dea}A. Deandrea et al., Preprint UGVA-DPT 1993/07-824.
\bibitem{rieckert}V. Rieckert, priv. communication.
\bibitem{cabb}N. Cabibbo, L. Maiani, \PL B79,109 (1978).
\bibitem{sharma} V. Sharma, DPF 94, Albuquerque (1994).
\bibitem{lepage} G.T. Bodwin, E. Bratten, T.C. Yuan, G.P. Lepage,
Phys. Rev. D46, 3703 (1992).
Glasgow, CLEO-CONF-94-11.
\bibitem{cleopsi} CLEO Collaboration, International Conference on HEP,
Glasgow, CLEO-CONF-94-11 (GLS 0248).
\end{thebibliography}
\end{document}